\documentclass[twocolumn,showpacs,showkeys]{revtex4}
\usepackage{graphicx}

\usepackage{psfrag}
\begin{document}

\title{Instantaneous Bethe-Salpeter Equation and Its Exact Solution}
\author{Chao-Hsi Chang$^{1,2}$, Jiao-Kai Chen$^{2,3}$,
Xue-Qian Li$^{1,4}$ and Guo-Li Wang$^{4,5}$}
\address{$^1$
CCAST (World Laboratory), P.O. Box 8730, Beijing 100080,
China\footnote{Not post-mail address.}\\
$^2$ Institute of Theoretical Physics, Chinese Academy of
Sciences, P.O. Box 2735, Beijing 100080, China\\
$^3$ Graduate School of the Chinese Academy of Sciences, Beijing
100039, China\\
$^4$ Department of Physics, NanKai University, TianJin 300071,
China\\
$^5$ Department of physics, FuJian Normal University, FuZhou
350007, China.}



\begin{abstract}
We present an approach to solve a Bethe-Salpeter (BS) equation
exactly without any approximation if the kernel of the BS equation
exactly is instantaneous, and take positronium as an example to
illustrate the general features of the solutions. As a middle
stage, a set of coupled and self-consistent integration equations
for a few scalar functions can be equivalently derived from the BS
equation always, which are solvable accurately. For positronium,
precise corrections to those of the Schr\"odinger equation in
order $v$ (relative velocity) in eigenfunctions, in order $v^2$ in
eigenvalues, and the possible mixing, such as that between $S$
($P$) and $D$ ($F$) components in $J^{PC}=1^{--}$
($J^{PC}=2^{++}$) states as well, are determined quantitatively.
Moreover, we also point out that there is a problematic step in
the classical derivation which was proposed first by E.E.
Salpeter. Finally, we emphasize that for the effective theories
(such as NRQED and NRQCD etc) we should pay great attention on the
corrections indicated by the exact solutions.

\end{abstract}


 \maketitle

Bethe-Salpeter (BS) equation \cite{BS} is a very good tool to
treat various bound state systems. For a fermion-antifermion
binding system, the BS equation is written as follows,
\begin{equation}
(\not\!{p_{1}}-m_{1})\chi_{P}(q)(\not\!{p_{2}}+m_{2})=
i\int\frac{d^{4}k}{(2\pi)^{4}}V(P,q,k)\chi_{P}(k)\;, \label{eq1}
\end{equation}
where $\chi_{P}(q)$ is the BS wave function, $P$ is the total
momentum, $q$ is relative momentum, and $V(P,k,q)$ is the
so-called BS kernel between the electron and positron in the bound
state, $p_{1}, p_{2}$ are the momenta of the constituent electron
$1$ and positron $2$, respectively. The total momentum $P$ and the
relative momentum $q$ are related to the momenta of the two quarks
as follows:
$p_{1}={\alpha}_{1}P+q, \;\;
{\alpha}_{1}=\frac{m_{1}}{m_{1}+m_{2}},\;\; p_{2}={\alpha}_{2}P-q,
\;\; {\alpha}_{2}=\frac{m_{2}}{m_{1}+m_{2}}.$

If the kernel of the four-dimensional BS equation, $V(P,k,q)$, in
$\vec{P}=0$ frame (C.M.S) of the concerned bound state, has the
behavior: $V(P,q,k)|_{\vec{P}=0}=V(\vec{q},\vec{k})\,,$ the BS
equation is called as `instantaneous one', and may be derived to a
Schr\"odinger equation accordingly, that was firstly realized by
E.E. Salpeter  \cite{salp}.

In Coulomb gauge, the transverse photon exchange is considered as
higher order, so at the lowest order the BS equation for
positronium has the kernel $V(P,q,k)|_{postronium}=\gamma^0 V_v
\gamma^0=-\gamma^0
\frac{4\pi\alpha}{\left(\vec{q}-\vec{k}\right)^{2}}\gamma^0$ only,
that is instantaneous, so we take positronium as an example to
pursue exact solutions of the BS equation. It certainly is
interesting that the relativistic corrections, including the
possible mixing, such as that between $S$ and $D$ components in
$J^{PC}=1^{--}$ state and that between $P$ and $F$ components in
$J^{PC}=2^{++}$ state etc, will be fully fixed by the
instantaneous BS equation. Here we report that as the case of
positronium, an instantaneous BS equation can be really solved
without any approximation, and outline the approach briefly. We
derive the instantaneous BS equation (the BS wave function in
$4\time 4$ spin structure) into a set of coupled and
self-consistent integration equations for its components (scalar
functions) without any approximation, and solve them numerically.
The accuracy of the solutions can be contral at all. The results
are discussed finally. While the details are put in
Refs.\cite{changw1,changw2}.

The approach which we show here mainly is to follow E.E. Salpeter
derivation but without any approximation.

As done in Ref.\cite{salp} and in book \cite{itz}, the
instantaneous BS wave function $\varphi_{P}(\vec{q})$ is
introduce:
\begin{equation}
\varphi_{P}(\vec{q})\equiv i\int
\frac{dq^0}{2\pi}\chi_{P}(q^0,\vec{q})\,, \label{eqinstw}
\end{equation}
then the BS equation Eq.\ref{eq1} is re-written
\begin{equation}
\label{eqBSin} \chi_{P}(q^0,\vec{q})=S_f^{(1)}(p^\mu_{1})
\eta(\vec{q})S_f^{(2)}(-p^\mu_{2})\,,
\end{equation}
where $S_f^{(1)}(p_{1})$ and $S_f^{(2)}(-p_{2})$ are the
propagators of the fermion and anti-fermion respectively and the
integrated `BS-nut'
$\eta(\vec{q})\equiv\int\frac{d^3\vec{k}}{(2\pi)}
V(\vec{q},\vec{k})\varphi_{P}(\vec{k})\,.$ For general
applications,
we keep $m_1\neq m_2$ at this moment, although final application
in this paper is to positronium $m_1=m_2$. The propagators can be
decomposed as:
\begin{eqnarray}
& -iJS_f^{(i)}(Jp^\mu_{i})=\frac{\Lambda^{+}_{i}(\vec{q})}{Jq^0
+\alpha_{i}M-\omega_i+i\epsilon}+
\frac{\Lambda^{-}_{i}(\vec{q})}{Jq^0+\alpha_{i}M+
\omega_{i}-i\epsilon}\,, \label{eqpropo}
\end{eqnarray}
with $\omega_{i}=\sqrt{m_{i}^{2}+\vec{q}^{2}}\,,\;\;\;\;
\Lambda^{\pm}_{i}(\vec{q})= \frac{1}{2\omega_{i}}\Big[
\gamma^0\omega_{i}\pm J(m_{i} +\vec{\gamma}\cdot\vec{q})\Big]\,,$
where $J=1$ for the quark($i=1$) and $J=-1$ for the
anti-quark($i=2$). It is easy to check
\begin{eqnarray}
&\Lambda^{\pm}_{i}(\vec{q})+ \Lambda^\mp_{i}(\vec{q})=\gamma^0\,,
\nonumber \\
&\Lambda^\pm_{i}(\vec{q}) \gamma^0
\Lambda^\mp_{i}(\vec{q})=0\,,\;\;\;\;
\Lambda^\pm_{i}(\vec{q})\gamma^0\Lambda^\pm_{i}(\vec{q})
=\Lambda^\pm_{i}(\vec{q})\,,\label{eqpropo1}
\end{eqnarray} Namely
$\Lambda^{\pm}$ are `energy' projection operators and complete.
With them for below discussions let us define
$\varphi^{\pm\pm}_{P}(\vec{q})$ as:
\begin{equation} \label{defini}
\varphi_P^{\pm\pm}(\vec{q})\equiv \Lambda^{\pm}_{1}(\vec{q})
\gamma^0\varphi_{P}(\vec{q})\gamma^0 \Lambda^{\pm}_{2}(\vec{q})\,.
\end{equation}
Because of the completeness of the projectors $\Lambda^{\pm}$, for
the BS wave function $\varphi_{P}(\vec{q})$, we have:
$\varphi_{P}(\vec{q})=\varphi^{++}_{P}(\vec{q})
+\varphi^{+-}_{P}(\vec{q})+\varphi^{-+}_{P}(\vec{q})
+\varphi^{--}_{P}(\vec{q})\,.$
As done by E.E. Salpeter, to derive the instantaneous BS equation
to corresponding Schr\"odinger equation, by carrying out a contour
integration of the time-component $q^0$ on both sides of
Eq.(\ref{eqBSin}), we obtain
\begin{eqnarray}
\label{eqcont} &
\varphi_{P}(\vec{q})=\frac{\Lambda^{+}_{1}(\vec{q})
\eta_{P}(\vec{q})\Lambda^{+}_{2}(\vec{q})}
{(M-\omega_{1}-\omega_{2})} -\frac{\Lambda^{-}_{1}(\vec{q})
\eta_{P}(\vec{q})\Lambda^{-}_{2}(\vec{q})}
{(M+\omega_{1}+\omega_{2})}\,,
\end{eqnarray}
and applying the complete set of the projection operators
$\Lambda^\pm_{iP}(\vec{q})$ to Eq.(\ref{eqcont}) further, we
obtain the coupled equations:
\begin{eqnarray}
\label{eqpp} &(M-\omega_{1}-\omega_{2})\varphi^{++}_{P}(\vec{q})=
\Lambda^{+}_{1}(\vec{q})
\eta_{P}(\vec{q})\Lambda^{+}_{2}(\vec{q})\,,
\end{eqnarray}
\begin{eqnarray}
\label{eqmm} &(M+\omega_{1}+\omega_{2})\varphi^{--}_{P}(\vec{q})=
-\Lambda^{-}_{1}(\vec{q})
\eta_{P}(\vec{q})\Lambda^{-}_{2}(\vec{q})\,,
\end{eqnarray}
\begin{eqnarray}
\label{eqpm}
\varphi^{+-}_{P}(\vec{q})=\varphi^{-+}_{P}(\vec{q})=0\,.
\end{eqnarray}
Note that because of the completeness of the projector set,
Eqs.(\ref{eqpp}, \ref{eqmm}, \ref{eqpm}) in whole are exactly
equivalent to Eq.(\ref{eqcont}) and Eq.(\ref{eqBSin}).

While E.E. Salpeter \cite{salp} and the authors of \cite{itz}
would like to connect the above coupled equations to the Breit
equation \cite{breit,breit1}, so they `combined'
Eqs.(\ref{eqpp},\ref{eqmm},\ref{eqpm}) into one operator equation
(in C.M.S. of the bound state):
\begin{eqnarray}
&[M-H_1(\vec{q})-H_2(\vec{q})]\varphi(\vec{q}) \nonumber \\
&= \Lambda^{+}_{1}(\vec{q})\gamma^0\eta(\vec{q})\gamma^0
\Lambda^{+}_{2}(\vec{q})-\Lambda^{-}_{1}(\vec{q})
\gamma^0\eta(\vec{q})\gamma^0\Lambda^{-}_{2}(\vec{q})\,,
\label{combi}
\end{eqnarray}
with the definitions $H_1(\vec{q})\equiv
m_1\beta+\vec{q}\cdot\vec{\alpha}\,, H_2(\vec{q})\equiv
m_2\beta-\vec{q}\cdot\vec{\alpha}\,,$ $\beta=\gamma^0\,,
\vec{\alpha}=\beta\vec{\gamma}$. Namely they considered the
equation Eq.(\ref{combi}) with the definition $H_{1,2}$ as an
operator representation of the coupled equations Eqs.(\ref{eqpp},
\ref{eqmm}, \ref{eqpm}). {\bf In fact, it is no correct.}
Eq.(\ref{combi}) is not fully equivalent to the coupled-equations
Eqs.(\ref{eqpp}, \ref{eqmm}, \ref{eqpm}).

Now let us show the un-equivalence as follows.

When applying the project operator
$\Lambda^{+}_1(\vec{q})\gamma^0\otimes\gamma^0\Lambda^{+}_{2}(\vec{q})$
to Eq.(\ref{combi}) we obtain Eq.(\ref{eqpp}), when applying the
project operator
$\Lambda^{-}_1(\vec{q})\gamma^0\otimes\gamma^0\Lambda^{-}_{2}(\vec{q})$
to Eq.(\ref{combi}) we obtain the Eq.(\ref{eqmm}), whereas when
applying
$\Lambda^{\pm}_1(\vec{q})\gamma^0\otimes\gamma^0\Lambda^{\mp}_{2}(\vec{q})$
to Eq.(\ref{combi}), then we obtain:
\begin{eqnarray}
&[M-\omega_1(\vec{q})+\omega_2(\vec{q})]\varphi^{+-}(\vec{q})=0\,,
\nonumber\\
&[M+ \omega_1(\vec{q})
-\omega_2(\vec{q})]\varphi^{-+}(\vec{q})=0\,, \label{eqpm11}
\end{eqnarray}
with $\omega_{1,2}=\sqrt{m^2_{1,2}+\vec{q}^2}$. When $m_1=m_2$
i.e. $\omega_1=\omega_2$, we may obtain Eq.(\ref{eqpm}) from
Eq.(\ref{eqpm11}) unless $M=0$ in the extremely relativistic
cases, thus in the cases $m_1=m_2$, equations Eq.(\ref{eqpm11})
are equivalent to Eq.(\ref{eqpm}). Whereas when $m_1\neq m_2$ i.e.
$\omega_1\neq \omega_2$, the equations Eq.(\ref{eqpm11}) have the
`trivial solutions' just as Eq.(\ref{eqpm}), but also have
`non-trivial solutions'. Practically when solving equations, such
as Eq.(\ref{combi}), one would not check the projections
$\Lambda^{+}_1(\vec{q})\gamma^0\otimes\gamma^0\Lambda^{+}_{2}(\vec{q})$,
$\Lambda^{-}_1(\vec{q})\gamma^0\otimes\gamma^0\Lambda^{-}_{2}(\vec{q})$
and
$\Lambda^{\pm}_1(\vec{q})\gamma^0\otimes\gamma^0\Lambda^{\mp}_{2}(\vec{q})$
on the solutions precisely, and generally non-trivial solutions
are chosen. Especially, when further approximations are made,
certain misleading may occur i.e. the obtained solutions may not
satisfy the equations Eq.(\ref{eqpm}), even in the cases $m_1=m_2$
\cite{changw1}. Therefore, we conclude that Eq.(\ref{eqpm11}) are
not fully equivalent to Eq.(\ref{eqpm}).

We will present the different consequences due to Eq.(\ref{eqpm})
and Eq.(\ref{eqpm11}) more precisely and solve them respectively
in Ref.\cite{changw3}.

In the literature the authors of Refs.\cite{salp,itz} etc made an
additional assumption on the spin structure of the BS wave
function e.g. the spin structure of the wave functions for
$S$-wave positronium has the formulation (in C.M.S. $\vec{P}=0$):
\begin{eqnarray}
&\varphi_{0^{-+}(^{1}S_0)}(\stackrel{\rightarrow}{q})=
\Big[{\gamma_0}+1\Big]\gamma_5 f(\stackrel{\rightarrow}{q})
\label{00-+}
\end{eqnarray}
for $J^{PC}=0^{-+}\, (^1S_0)$; and
\begin{eqnarray}
&\varphi_{1^{--}(^{3}S_1),\lambda}(\stackrel{\rightarrow}{q})=
\Big[{\gamma_0}+1\Big]A_\lambda f(\stackrel{\rightarrow}{q})
\label{01--}
\end{eqnarray}
for $J^{PC}=1^{--}\,(^3S_1)$. Here $A_\lambda\equiv
\epsilon_\lambda^{+}\gamma^{-}+\epsilon_\lambda^{-}\gamma^{+}
-\epsilon_\lambda^{\Delta}\gamma^{\Delta}$. With the spin
structure finally the bound state problem turned to solve a
Schr\"odinger equation accordingly by taking the so-called
positive-energy equation Eq.(\ref{eqpp}) only (the rests are
ignored).

To solve the instantaneous BS equation Eq.(\ref{eqBSin}) for
positronium without any approximation, even in the case $m_1=m_2$
we `return' to solve the Eqs.(\ref{eqpp}, \ref{eqmm}, \ref{eqpm}).
In fact, the approach here is also applicable in the cases
$m_1\neq m_2$.

In $\vec{P}=0$ frame (C.M.S.), the most general formulation of the
BS wave function for the bound state $J^{PC}=0^{-+}$ ($^1S_0$) may
be written as the follows:
\begin{eqnarray}
\label{gg0-+}
&\displaystyle\varphi_{^{1}S_0}(\stackrel{\rightarrow}{q})
=\gamma^{0}\gamma^{5}\varphi_{1}(|\vec{q}|)+\gamma^{5}
\varphi_{2}(|\vec{q}|)\nonumber\\
&\displaystyle +
\sqrt{\frac{4\pi}{3}}\gamma^{5}\frac{|\vec{q}|}{2m}E\varphi_{3}(|\vec{q}|)
\gamma^{0}
+\sqrt{\frac{4\pi}{3}}\gamma^{5}\frac{|\vec{q}|}{2m}E\varphi_{4}(|\vec{q}|)\,,
\end{eqnarray}
where $E\equiv \Big[Y_{1-1}\gamma^{+}+ Y_{11}\gamma^{-}-
Y_{10}\gamma^{3}\Big]\,,$
$\gamma^{+}\equiv-\frac{\gamma^{1}+i\gamma^{2}}{\sqrt{2}}\,,$
$\gamma^{-}=\frac{\gamma^{1}-i\gamma^{2}}{\sqrt{2}}$ and
$Y_{lm}\equiv Y_{lm}(\theta_q,\phi_q)$ are spherical harmonics. To
apply the equation Eq.(\ref{eqpm}) to Eq.(\ref{gg0-+}), we obtain
the `constraints' for the components of the wave function
precisely:
\begin{eqnarray}
\varphi_1(|\vec{q}|)= -\frac{1}{2}\varphi_4(|\vec{q}|)\;, \;\;\;
\varphi_3(|\vec{q}|)=0\;. \label{constrai1}
\end{eqnarray}
$\varphi_3, \varphi_4$ in Eq.(\ref{gg0-+}) are replaced with
Eq.(\ref{constrai1}), then the relativistic wave function of the
state $0^{-+}$ contain two independent components $\varphi_1,
\varphi_2$ only. With straightforward calculations, from
Eqs.(\ref{eqpp}, \ref{eqmm}) for positronium, finally we obtain
the two coupled radius equations:
\begin{eqnarray}
&(M-2m)f_{1}+2mf_{1}-2mf_{2}=\nonumber\\
&-\frac{\alpha_{s}}{\pi}
\frac{m}{\omega}\int\frac{|\vec{k}|}{|\vec{q}|}d|\vec{k}|
Q_{0}f_{2}\,, \nonumber\\
&(M-2m)f_{2}-2\frac{\omega^{2}}{m}f_{1}+2mf_{2}=\nonumber\\
&-\frac{\alpha_{s}}{\pi}\frac{m}{\omega}
\int\frac{|\vec{k}|}{|\vec{q}|}d|\vec{k}|
\left(Q_{0}+\frac{|\vec{q}||\vec{k}|}{m^2}Q_{1}\right)f_{1}\,,
\label{compeqr2}
\end{eqnarray}
where $Q_{n}\equiv
Q_{n}(\frac{|\vec{q}|^{2}+|\vec{k}|^{2}}{2|\vec{q}||\vec{k}|})\;\;(n=0,1,
\cdots)$ are the $n$ the Legendre functions of the second kind.
Here to shorten the notations we have introduced $f_{1,2}$:
$\varphi_{1,2}(|\vec{q}|)\equiv f_{1,2}$ and
$\omega=\omega_{1,2}$, and the angular integrations have been
carried out precisely. Now without any approximation a
self-consistent problem, two coupled equations Eq.(\ref{compeqr2})
for two independent scalar functions $f_{1,2}$, is reached to.

Then we further solve the coupled equations Eq.(\ref{compeqr2})
numerically by expanding the components $f_{1,2}$ and the
equations in terms of the bases of the exact solutions of the
Schr\"odinger equation \cite{onetwo}, and then diagonalizing the
matrix equations.

The final results for the eigenvalues $E^{(j)},\;\;j=1, 2,
\cdots$:
\begin{eqnarray}
&E^{(1)}=-6.8026952534\,;\;\;\;\;\;\;
E^{(2)}=-1.70069524809\,;\nonumber\\
&E^{(3)}=-0.75586715480\,;\;\;\;\; E^{(4)}=-0.42517586843\,;
\cdots \label{0-+val}
\end{eqnarray}
and the corresponding eigenfunctions
\begin{eqnarray}
f^{(j)}_i(|\vec{q}|)=\sum_{nl} C^{(j)}_{i,nl}\cdot
R_{nl}(|\vec{q}|)\,, \label{solution}
\end{eqnarray}
where $(j)$ denotes those correspond to the $j$th eigenvalue
Eq.(\ref{0-+val}), $R_{nl}(|\vec{q}|)$ ($nl=1S,\, 2S,\, 2P,\,
3S,\, 3P,\, \cdots$) are the Schr\"odinger radius solutions in
momentum representation for the positronium and their precise
expression can be found in Ref.\cite{onetwo}. For the low-laying
solutions the coefficients $C^{(j)}_{i,nl}\; (nl\leq 4S)$ are put
in TABLE I and $C^{(j)}_{i,nl}\leq 10^{-6}\; (nl > 4S)$.
\begin{table}
\begin{center}
\caption{The expansion coefficients $C^{(j)}_{i,nl}$ for the
$J^{PC}=0^{-+}$ states.} \vspace{2mm}
\begin{tabular}{|r|c|c|c|c|c|c|}
\hline\hline  & NL & WF &$ C^{(j)}_{i,1S}$ & $ C^{(j)}_{i,2S}$ & $
C^{(j)}_{i,3S}$ & $C^{(j)}_{i,4S}$ \\
\hline  & $E^{(1)}$ & $f^{(1)}_{1}$&-0.707104 & 8.934D-6 &
4.225D-6 & 2.629D-6 \\
\cline {3-7} {}  & {($1^1S_0$)} &$f^{(1)}_{2}$&-0.707109 &
6.962D-6
& 3.206D-6 &1.979D-6\\
\cline{2-7} {} & $E^{(2)}$ &$f^{(2)}_{1}$& 5.031D-6 &
0.707106 & -7.246D-6 & -3.571D-6  \\
\cline{3-7} {} & {($2^1S_0$)} &$f^{(2)}_{2}$ & 7.004D-6 & 0.707107
&-6.567D-6 &-3.169D-6 \\
\cline{2-7} {} & $E^{(3)}$ &$f^{(3)}_{1}$& 2.188D-6 &
5.898D-6 & 0.707107 & -6.768D-6  \\
\cline{3-7} {}  & {($3^1S_0$)} &$f^{(3)}_{2}$ & 3.207D-6 &
6.577D-6 & 0.707107 &
-6.425D-6  \\
\hline\hline
\end{tabular}
\end{center}
\end{table}
Based on the values in TABLE I, up to the order $v$ accuracy the
eigenfunctions for $J^{PC}=0^{-+}(^1S_0)$ states, in fact, can be
simplified to write as
\begin{eqnarray}
\varphi_{^{1}S_0}(\stackrel{\rightarrow}{q})\simeq
\Bigg\{{\gamma^0}+1+\sqrt{\frac{4\pi}{3}}E\gamma^{0}
\frac{|\vec{q}|}{m}\Bigg\}f\gamma^5 \label{a00-+}
\end{eqnarray}
with $f_1(\stackrel{\rightarrow}{q})\simeq
f_2(\stackrel{\rightarrow}{q})\equiv f=R_{n(l=0)}(|\vec{q}|)$.

Let us now solve the equations for the $1^{--}$ states.

Similar to $0^{-+}$ states, we start with the most general BS wave
function, and have the constraints Eq.(\ref{eqpm}) applied, then
the BS wave function becomes the form:
\begin{eqnarray}
&\varphi_{1^{--},\lambda}(\stackrel{\rightarrow}{q})=
\Bigg(2\sqrt{\frac{\pi}{3}}C_\lambda+A_\lambda\Bigg)f_{1}
\nonumber\\
&+\Bigg(-2\sqrt{\frac{\pi}{3}}D_\lambda\gamma^{5}
+A_\lambda\gamma^{0}\Bigg)f_{2}
+\Bigg(-2\sqrt{\frac{5}{3}}C_\lambda
+B_\lambda\Bigg)f_{3}\nonumber \\
& +\Bigg(\sqrt{\frac{5}{3}}D_\lambda\gamma^{5}
-B_\lambda\gamma^{0}\Bigg)f_{4} \,,\label{g1--r}
\end{eqnarray}
where $A_\lambda$ is the same as in Eq.(\ref{01--}) and
\begin{eqnarray}
&B_\lambda\equiv\sqrt{6}\epsilon_\lambda^{+}\gamma^{+}Y_{2-2}-
\sqrt{3}\Big(\epsilon_\lambda^{+}\gamma^{\Delta}
+\epsilon_\lambda^{\Delta}\gamma^{+}\Big)Y_{2-1}\nonumber\\
&+\Big(\epsilon_\lambda^{+}\gamma^{-}+\epsilon_\lambda^{-}\gamma^{+}
+2\epsilon_\lambda^{\Delta}\gamma^{\Delta}\Big)Y_{20}\nonumber\\
& -\sqrt{3}\Big(\epsilon_\lambda^{-}\gamma^{\Delta}
+\epsilon_\lambda^{\Delta}\gamma^{-}\Big)Y_{21}
+\sqrt{6}\epsilon_\lambda^{-}\gamma^{-}Y_{22}\,,\nonumber\\
&C_\lambda\equiv
\frac{|\vec{q}|}{m}\Big[\epsilon_\lambda^{+}Y_{1-1}+\epsilon_\lambda^{-}Y_{11}
-\epsilon_\lambda^{\Delta}Y_{10}\Big]
\,, \nonumber\\
&D_\lambda\equiv
\frac{|\vec{q}|}{m}\Big[\Big(\epsilon_\lambda^{-}\gamma^{\Delta}
-\epsilon_\lambda^{\Delta}\gamma^{-}\Big)Y_{11}
+\Big(\epsilon_\lambda^{\Delta}\gamma^{+}
-\epsilon_\lambda^{+}\gamma^{\Delta}\Big)Y_{1-1}\nonumber\\
& +\Big(\epsilon_\lambda^{+}\gamma^{-}
-\epsilon_\lambda^{-}\gamma^{+}\Big)Y_{10}\Big]\,.\nonumber
\end{eqnarray}
Here $\Delta\equiv 0$ is introduced for the $0^{th}$ component to
avoid confusions. From Eq.(\ref{g1--r}), we may see clearly that
$f_{3,4}$ correspond to $D$-wave components while $f_{1,2}$ to
$S$-wave ones. The wave function now has four independent
components $f_i (i=1,\dots,4)$.

From Eqs.(\ref{eqpp},\ref{eqmm}) and to carry out the angular
integration in the equations, finally we obtain the coupled radius
equations:
\begin{eqnarray}
&(M-2m)f_1=\frac{2\sqrt{5}}{3\sqrt{\pi}}
\frac{|\vec{q}|^{2}}{m}f_{4}-2mf_1
-\frac{2m^{2}+4\omega^{2}}{3m}f_2
\nonumber  \\
& -\frac{\alpha}{\pi}\int\frac{|\vec{k}|}{|\vec{q}|}d|\vec{k}|
\Big[\sqrt{\frac{5}{9\pi}}
\frac{|\vec{q}||\vec{k}|}{m\omega}Q_{1}f_{4}
-\left(\frac{2}{3}\frac{|\vec{q}||\vec{k}|}{m\omega}Q_{1}
+\frac{m}{\omega}Q_{0}\right)f_2 \Big]\,,
\nonumber
\end{eqnarray}
\begin{eqnarray}
&(M-2m)f_2=\frac{2\sqrt{5}}{3\sqrt{\pi}}
\frac{|\vec{q}|^{2}}{m}f_{3}
-\frac{2}{3}\frac{2m^{2}+\omega^{2}}{m}f_1-2mf_2
\nonumber  \\
&-\frac{\alpha}{\pi}\int \frac{|\vec{k}|}{|\vec{q}|}d|\vec{k}|
\Big[\frac{1}{3}\sqrt{\frac{5}{\pi}}
\frac{|\vec{q}||\vec{k}|}{m\omega}Q_{1}f_{3}-\left(\frac{1}{3}
\frac{|\vec{q}||\vec{k}|}{m\omega}Q_{1}
+\frac{m}{\omega}Q_{0}\right)f_1 \Big]\,,
\nonumber
\end{eqnarray}
\begin{eqnarray}
&(M-2m)f_{3}=-2mf_{3}+\frac{2}{3}\frac{2m^{2} +\omega^{2}}{m}f_{4}
-\frac{4\sqrt{\pi}}{3\sqrt{5}}\frac{|\vec{q}|^{2}}{m}f_2
\nonumber  \\
&-\frac{\alpha}{\pi}\int \frac{|\vec{k}|}{|\vec{q}|}d|\vec{k}|
\Big[\left(\frac{|\vec{q}||\vec{k}|}{3m\omega}Q_{1}
+\frac{m}{\omega}Q_{2}\right)f_{4}
-\frac{2\sqrt{\pi}}{3\sqrt{5}}\frac{|\vec{q}||\vec{k}|}{m\omega}Q_{1}f_2\Big]\,,
\nonumber
\end{eqnarray}
\begin{eqnarray}
&(M-2m)f_{4}=\frac{1}{3}\frac{2m^{2} +4\omega^{2}}{m}f_{3}-2mf_{4}
-\frac{4\sqrt{\pi}}{3\sqrt{5}}\frac{|\vec{q}|^{2}}{m}f_1 \nonumber  \\
& -\frac{\alpha}{\pi}\int\frac{|\vec{k}|}{|\vec{q}|}d|\vec{k}|
\Big[\left(\frac{m}{\omega}Q_{2}
+\frac{2}{3}\frac{|\vec{q}||\vec{k}|}{m\omega}Q_{1}\right)f_{3}\nonumber  \\
&-\frac{2\sqrt{\pi}}{3\sqrt{5}}\frac{|\vec{q}||\vec{k}|}{m\omega}Q_{1}f_1\Big]\,.
\label{finr-1--4}
\end{eqnarray}
Four coupled equations for four independent components
$f_{1,2,3,4}$ is the request of the $S$-$D$ wave mixing and a
self-consistent problem is reached.

We solve the Eqs.(\ref{finr-1--4}) numerically with the same
method as adopted in the case for $0^{-+}$, and obtain the
results:
\begin{eqnarray}
&E^{(1)}(1^3S_1)=-6.8027275\,;\;\;
E^{(2)}(2^3S_1)=-1.7007008\,; \nonumber\\
&E^{(3)}(3^3S_1)=-0.7558690\,;\;\;
E^{(4)}(3^3D_1)=-0.7558729\,;\nonumber\\
&\cdots \label{e-lev1}
\end{eqnarray}
for eigenvalues, while the table of the coefficients
$C^{(j)}_{i,nl}$ for the eigenfunctions is too big to present
here. Instead, we present them here only up to the order $v$
accuracy as in the case for $0^{-+}$.

For $J^{PC}=1^{--} (^3S_1)$ states we can simplify:
$f_1(\stackrel{\rightarrow}{q}) \simeq
-f_2(\stackrel{\rightarrow}{q})\equiv f=R_{n(l=0)}(|\vec{q}|)$ and
$f_3(\stackrel{\rightarrow}{q})\simeq
f_4(\stackrel{\rightarrow}{q})\simeq 0$, i.e.
\begin{eqnarray}
&\varphi_{^{3}S_1,\lambda}(\stackrel{\rightarrow}{q})
\simeq\Big[\Big(2\sqrt{\frac{\pi}{3}}\frac{|\vec{q}|}{m}C_\lambda+A_\lambda\Big)
\nonumber\\
&-\Big(-2\sqrt{\frac{\pi}{3}}\frac{|\vec{q}|}{m}D_\lambda\gamma^{5}
+A_\lambda\gamma^{0}\Big)\Big]f\,. \label{a101--}
\end{eqnarray}
For $J^{PC}=1^{--} (^3D_1)$ states we can simplify:
$f_1(\stackrel{\rightarrow}{q}) \simeq
f_2(\stackrel{\rightarrow}{q})\simeq 0$ and
$f_3(\stackrel{\rightarrow}{q})\simeq
f_4(\stackrel{\rightarrow}{q})\equiv f=R_{n(l=2)}(|\vec{q}|)$,
i.e.
\begin{eqnarray}
&\varphi_{^{3}D_1,\lambda}(\stackrel{\rightarrow}{q})\simeq
\Big[\Big(-2\sqrt{\frac{5}{3}}\frac{|\vec{q}|}{m}C_\lambda+B_\lambda\Big)
\nonumber\\
&+\Big(\sqrt{\frac{5}{3}}\frac{|\vec{q}|}{m}D_\lambda\gamma^{5}
-B_\lambda\gamma^{0}\Big)\Big]f \label{a201--}
\end{eqnarray}
The accurate coefficients $C^{(j)}_{i,nl}$ for $J^{PC}=1^{--}$ can
be found in our achieved paper \cite{changw2}.

We may see that the exact solutions obtained here for the wave
function contain the order $v=\frac{|\vec{q}|}{m}$ corrections to
the approximate ones obtained by E.E. Salpeter approaches
(described as above), namely, the terms in Eqs.(\ref{a00-+},
\ref{a101--}, \ref{a201--}) explicitly proportional to $v$. The
corrections will cause the normalization for the wave function a
change in order $v^2$ (the detail in \cite{changw2}). The most
interesting fact is the exact solutions for $1^{--}$ states
present the possible $S-D$ wave mixing. To compare the eigenvalues
Eq.(\ref{0-+val}) for $0^{-+}$ and Eq.(\ref{e-lev1}) for $1^{--}$
with those of the Schr\"odinger solution (the degenerate ones
without any splitting), we may see the `hyperfine' splitting for
spin singlet and triplet states which is in order of $\alpha^2$
($v \simeq \alpha$), and there is a remarkable splitting between
$n[^3S_1]$ and $n[^3D_1] (n\geq 3)$ for $1^{--}$ states
$E^{(3)}(3^3S_1)\geq E^{Sch}_3\geq E^{(4)}(3^3D_1)$ and $\Delta
E\simeq \alpha^2$ due to the fact that the BS equation
Eq.(\ref{eqBSin}) is relativistic.

Note that the approach we present here may be applicable for all
the instantaneous BS equations and can obtain the exact solutions,
while the features of the exact solutions for positronium
described above may be different for an instantaneous BS equation
with a different kernel from that of Coulomb interaction.

Based on a specific problem of positronium, we have learnt that
the wave functions contain quite great relativistic corrections
thus for effective theories, such as NRQED \cite{nrqed} (and NRQCD
\cite{braat}), the relativistic corrections from the wave
functions should be considered carefully if the relevant
calculations declare the relativistic effects are taken into
account. For NRQED (NRQCD) etc, the wave function effects may be
involved either at matching the underlined theory to the effective
one or into the non-pertubative matrix elements. Moreover the
problems, such as gauge invariance for the matrix elements and
$E1$ (electrical dipole) radiative transitions etc, are sensitive
to the effects, the exact solutions are crucial \cite{changw1}.

\noindent {\Large\bf Acknowledgement}: The author
(C.H. Chang) would like to thank Stephen L. Adler for valuable
discussions and to thank Xiangdong Ji for very useful discussions.
The authors would like to thank Tso-Hsiu Ho and Cheng-Rui Ching
for valuable suggestion and encouragement. This work was supported
in part by the National Natural Science Foundation of China.


\begin{thebibliography}{s2}


\bibitem{BS}
E.E. Salpeter and H.A. Bethe, {\it Phys. Rev.} {\bf 84}, 1232
(1951).

\bibitem{salp}
E.E. Salpeter, {\it Phys. Rev.} {\bf 87}, 328 (1952).

\bibitem{itz} C. Itzykson and J. Zuber, {\bf Quantum Field
Theory}, McGRAW-HILL Int. Book Company.

\bibitem{breit} G. Breit, Phys. Rev. {\bf 34}, 553 (1929).

\bibitem{breit1}
G. Breit and G.E. Brown, Phys. Rev. {\bf 74}, 1278 (1948); G.E.
Brown and D.G. Ravenhall, Proc. Roy. Soc. (London), {\bf A208},
552 (1951).

\bibitem{changw1}
Chao-Hsi Chang, Jiao-Kai Chen and Guo-Li Wang, hep-th/0312250.

\bibitem{changw2}
Chao-Hsi Chang, Jiao-Kai Chen, Xue-Qian Li and Guo-Li Wang, {\it
in preparation}.

\bibitem{changw3}
Chao-Hsi Chang, Jiao-Kai Chen, Xue-Qian Li and Guo-Li Wang, {\it
in preparation}.

\bibitem{onetwo} H.A. Bethe and E.E. Salpeter, {\it Quantum
Mechanics of One- and Two-Electron Atoms}, Springer-Verlag, 1957.

\bibitem{nrqed} W.E. Caswell and G.P. Lepage, Phys. lett.
{\bf 167}, 437 (1986).

\bibitem{braat} Geoffrey T. Bodwin, Eric Braaten and G. Peter
Lepage,  Phys. Rev. D {\bf 51}, 1125 (1995); Erratum ibid {\bf
55}, 5835 (1997).

\end{thebibliography}
\end{document}